\documentclass[cite]{epl}

\usepackage{amsmath}
\newcommand{\be}{\begin{equation}}
\newcommand{\ee}{\end{equation}}
\newcommand{\bea}{\begin{eqnarray}}
\newcommand{\eea}{\end{eqnarray}}
\newcommand{\beas}{\begin{eqnarray*}}
\newcommand{\eeas}{\end{eqnarray*}}
\newcommand{\la}{\label}

\newcommand{\p}{\partial}
\newcommand{\Order}{\mathrm{O}}
\newcommand{\order}{\mathrm{o}}
\newcommand{\e}{\mathrm{e}}
\newcommand{\ie}{{\it i.e. }}

\newcommand{\etalii}{{\it et al. }}
\newcommand{\avg}[1]{\left\langle #1 \right\rangle} 
\newcommand{\Tr}{\mathrm{Tr}}

\renewcommand{\b}[1]{{\bm{#1}}}  
\renewcommand{\r}{{\bm{r}}}

\newcommand{\A}{{\bm{A}}}
\newcommand{\br}{{\bm{r}}} 
\newcommand{\bk}{{\bm{k}}} 

\newcommand{\Fr}{{F^{\text{R}}}}

\title{The Casimir force at high temperature}
\author{P. R. Buenzli\thanks{E-mail: \email{Pascal.Buenzli@epfl.ch}} \and Ph. A. Martin}

\institute{                    
  Institute of Theoretical Physics - Swiss Federal Institute of Technology Lausanne,
  CH-1015 Lausanne EPFL, Switzerland
}

\begin{document}

\maketitle

\begin{abstract}
    The standard expression of the high-temperature Casimir force between perfect conductors is
    obtained by imposing macroscopic boundary conditions on the electromagnetic field at
    metallic interfaces. This force is twice larger than that computed in microscopic classical
    models allowing for charge fluctuations inside the conductors. We present a direct
    computation of the force between two quantum plasma slabs in the framework of non
    relativistic quantum electrodynamics including quantum and thermal fluctuations of both
    matter and field. In the semi-classical regime, the asymptotic force at large slab
    separation is identical to that found in the above purely classical models, which is
    therefore the right result.  We conclude that when calculating the Casimir force at
    non-zero temperature, fluctuations inside the conductors can not be ignored.
\end{abstract}

Casimir showed in 1948 \cite{casimir} that the zero-point energy of the quantum electromagnetic
field generates an attractive force between two perfectly conducting metallic plates at
distance $d$ and zero temperature. In his calculation, the microscopic structure of the
conductors is not taken into account. The latter are merely treated as macroscopic boundary
conditions for the electromagnetic fields requiring the vanishing of the tangential electric
field.  This geometrical constraint modifies the field eigenmodes depending on $d$. The
$d$-dependence of the modified zero-point energy is the source of the well known Casimir force
\be f^\text{vac}(d) = -\frac {\pi^2 \hbar c}{240\, d^4} \la{vac} \ee ($\hbar$ denotes Planck's
constant, $c$ the speed of light).

The generalisation of Casimir's calculation to thermalized fields was given some years later in
\cite{fierz, mehra}, see \cite{balian-duplantier} for a recent account.  When the temperature
$T$ is different from zero, one can form the dimensionless parameter $\alpha=\beta \pi \hbar
c/d$ (the ratio of the thermal wave length of the photon to the conductors separation; $\beta$
is the inverse temperature).  A large value of $\alpha$ (low temperature, short separation)
characterizes the quantum regime whereas a small value of $\alpha$ (high temperature, large
separation) yields a purely classical asymptotic result (independent of $\hbar$ and $c$)
\begin{align}
    f = -\frac{\zeta(3)}{4\pi\beta d^3} + \Order(\e^{-b/\alpha}), \quad
    \alpha\to 0, \quad b>0
\label{force-4pibeta}
\end{align}
where $\zeta(s)$ is the Riemann zeta function. Each field mode is a thermalized quantum
mechanical oscillator with frequencies obtained from the previously described macroscopic
boundary conditions. All fluctuations inside the conductors are ignored.  We note that in fact,
on purely dimensional grounds, a term $\propto d^{-3}$ must also be proportional to $k_B T$,
the only issue being the numerical value of the proportionality constant. This issue is the
subject of this letter.

In recent times, a number of works have adressed the question of the incidence of the
microscopic charge and field fluctuations inside the conductors on the Casimir force
\cite{forrester-janco-tellez}, \cite{janco-tellez}, \cite{buenzli-martin}. The considered
models are classical : the conductors are represented by slabs (or surfaces) containing mobile
charges in thermal equilibrium and interacting through the sole Coulomb potential. These models
all yield the same universal result for the mean electrostatic force between the slabs at fixed
temperature and large distance
\begin{align}
    \avg{f} = -\frac{\zeta(3)}{8\pi\beta d^3} + \order(d^{-3}), \quad
    d\to\infty \label{force-8pibeta}
\end{align}
Universality means that the asymptotic force does not involve any parameter characterizing the
material constitution of the conductors: particle charges and masses, densities $\rho$ and slab
thicknesses.\footnote{In microscopic conductor models, there is a new energy parameter
    $e^{2}/\rho^{-1/3}$, the mean potential energy, so that universality does not follow from a
    simple dimensional analysis.}  In \cite{forrester-janco-tellez}, the authors study a
statistical mechanical system of charges confined to a plane at distance $d$ of a macroscopic
(non fluctuating) planar conductor. In \cite{janco-tellez}, they show that replacing the above
macroscopic conductor by fluctuating charges does not alter the result
(\ref{force-8pibeta}). We provide in \cite{buenzli-martin} a general derivation of
(\ref{force-8pibeta}) showing that universality is guarantied by perfect screening sum rules
\cite{martin-sumrules}.

If one compares the result (\ref{force-8pibeta}) with (\ref{force-4pibeta}), one sees that the
extrapolation of Casimir's calculation to the classical regime is larger by a factor $2$ than
that obtained in the classical microscopic models. The two approaches are based on different
premises~: (\ref{force-4pibeta}) was derived from the frequency spectrum of the full
electromagnetic field but treating the metals as macroscopic bodies without internal
structure. One the contrary, the force in (\ref{force-8pibeta}) is purely electrostatic
(longitudinal field) and it originates from the particle fluctuations inside the conductors.

This calls for a more complete model that incorporates the dynamical part of the field
(transverse field) in addition to the internal degrees of freedom of the conductors. A
preliminary remark is in order: it is well known that classical matter in thermal equilibrium
always decouples from the transverse field because of the Bohr--van Leeuwen theorem
\cite{Alast}. It is therefore necessary to treat the conductors' charges quantum mechanically.
The complete model is formulated as follows.  One considers two parallel slabs $A$ and $B$ of
surface $L^{2}$, thickness $a$ and at a distance $d$ apart. They contain non relativistic
quantum charges (electrons, ions, nuclei) with appropriate statistics. The total charge in each
slab is taken equal to zero. The slabs are immersed in a quantum electromagnetic field, which
is itself enclosed into a larger box $K$ with sides of length $R,\,R\gg L, a$.  The Hamiltonian
of the total finite volume system reads in Gaussian units \footnote{The Pauli coupling terms
    between spins and magnetic field are not taken into account here.}
\begin{align}
    H=\sum_{i}\frac{\left({\b p}_{i} -
        \tfrac{e_{\gamma_{i}}}{c}\A(\r_{i})\right)^{2}}{2m_{\gamma_{i}}} + \sum_{i<j}
    \frac{e_{\gamma_{i}}e_{\gamma_{j}}}{|\r_{i}-\r_{j}|}+\sum_{i}V^{{\rm
            walls}}(\gamma_{i},\r_{i})+H_{0}^{{\rm rad}} \la{B.1}
\end{align}
The sums run on all particles with position $\r_{i}$ and species index $\gamma_{i}$; $V^{{\rm
        walls}}(\gamma_{i},\r_{i})$ is a steep external potential that confines the particles
in the slabs . It can eventually be taken infinitely steep at walls' position implying
Dirichlet boundary conditions for the particle wave functions.

The electromagnetic field is written in the Coulomb (or transverse) gauge so that the vector
potential $\A(\r)$ is divergence free and $H_{0}^{{\rm rad}}$ is the Hamiltonian of the free
radiation field.  For it we impose periodic boundary conditions on the faces of the large box
$K$.  Hence expanding $\A(\r)$ in the plane waves modes $\bk= (\tfrac{2\pi
    n_{x}}{R},\tfrac{2\pi n_{y}}{R},\tfrac{2\pi n_{z}}{R})$ gives the usual formulae
\begin{align}
	&\A(\r) = \left(\frac{4\pi \hbar c^{2}}{R^{3}}\right)^{1/2}\sum_{\bk,\lambda} g(\bk) \frac{
        \bm{e}_{\bk} (\lambda)}{\sqrt{2\omega_{\bk}}} (a_{\bk,\lambda}^{*} e^{-i\bk\cdot\r} +
    a_{\bk,\lambda}e^{i\bk\cdot\r}) \la{B.2}
    \\ &H_{0}^{\rm rad} = \sum_{\bk,\lambda} \hbar
    \omega_\bk\,a_{\bk,\lambda}^{*}a_{\bk,\lambda}, \qquad \omega_\bk = c |\bk|
\la{B.2a}
\end{align}
In (\ref{B.2}), $\bm{e}_{\bk}(\lambda)$, $\lambda=1,2$, are the polarization vectors and
$g(\bk), \; g(0)=1$, is a form factor that takes care of ultra-violet divergences.

We suppose that the matter in the slabs is in thermal equilibrium with the radiation field and
therefore introduce the finite volume free energy of the full system at temperature $T$
\begin{align}
	\Phi_{R, L, d}=-k_B T \ln \Tr e^{-\beta H} \label{free-energy}
\end{align}
where the trace $\Tr\equiv\Tr_\text{mat}\Tr_\text{rad}$ is carried over particles' and field's
degrees of freedom.  The force between the slabs by unit surface is now defined by
\begin{align}
	f(d) = \lim_{L\to\infty}\lim_{R\to\infty} f_{R,L}(d) \quad\text{with}\quad
    f_{R,L}(d) = -\frac{1}{L^2} \frac{\p}{\p d}  \Phi_{R, L, d} \label{fRL}
\end{align}
Adding and substracting the free energy of the free photon field in (\ref{free-energy}) leads
to
\begin{align}
	\Phi_{R, L, d}=-k_B T \ln \left( \frac{\Tr e^{-\beta H}}{Z_0^\text{rad}} \right) -
    k_B T \ln Z_0^\text{rad} \label{PhiRLd}
\end{align}
where $Z_0^\text{rad}$ is the partition function of the free photon field in the volume
$K$. Since the last term of (\ref{PhiRLd}) is independent of $d$, it does not contribute to the
force (\ref{fRL}). Therefore
\begin{align}
	f(d) = k_B T \lim_{L\to\infty}\lim_{R\to\infty} \frac{1}{L^2} \frac{\p}{\p d} \ln \left(
    \frac{\Tr e^{-\beta H}}{Z_0^\text{rad}} \right)
\label{force-def}
\end{align}
In principle $f(d)$ yields the Casimir force taking into account quantum and thermal
fluctuations of both matter and field.

The main result presented in this letter is that, in the semi-classical regime, the dominant
term of the large distance behaviour of the force (\ref{force-def}) is still given by the
universal classical behaviour (\ref{force-8pibeta}).  This regime is obtained when the particle
thermal wave lengths $\lambda_{\gamma}=\hbar(\beta/m_{\gamma})^{1/2}$ are much smaller than the
slabs' thickness and separation $(\lambda_{\gamma}\ll a,\;d)$.

More precisely, the force is of the form
\begin{align}
    f(d)=-\frac{\zeta(3)}{8\pi\beta d^{3}} +\mathcal{R}(\beta,\hbar,d),\quad\quad {\rm
        where}\quad \mathcal{R}(\beta,\hbar,d)=\Order(d^{-4}) \la{concl}
\end{align}
namely, the quantum corrections included in the remainder $\mathcal{R}(\beta,\hbar,d)$ only
occur at the subdominant order $d^{-4}$.

The formalism adapted to the investigation of the high temperature (or semi-classical) regime
is the Feynman-Kac-It\^o path integral representation of the Gibbs weight. In this formalism a
quantum point particle of species $\gamma$ is represented by a closed Brownian path
$\br+\lambda_\gamma\b\xi (s),\;0\leq s <1,\; \b\xi (0)=\b\xi (1)=0$, starting at $\br$ and of
extension $\lambda_\gamma$: it can be viewed as a charged random wire at $\r$. Thus the
ensemble of wires can be treated as a classical-like system with phase space points
$(\r_{i},\b\xi_{i})$. The wire shape $\lambda_\gamma\b\xi (s)$ (the quantum fluctuation) plays
the role of an internal degree of freedom; see \cite{BM}, section IV, for more details on this
formalism.  Here, for simplicity, we use Maxwell-Boltzmann statistics for the particles. We
also treat the field classically on the ground that the spacing between the dimensionless
energy levels $\beta \hbar \omega_\bk$ of the $\bk$ field mode become vanishingly small in the
high temperature and large distance asymptotics ($\alpha\ll1$). A complete presentation will be
found in \cite{sami}, \cite{buenzli-martin-loops}.

The Gibbs weight associated to $n$ wires is  
\begin{align}
    \exp\Biggl(-\beta\sum_{i<j}^{n}e_{\gamma_{i}}e_{\gamma_{j}}V(\r_{i},\b\xi_{i},\r_{j},\b\xi_{j})+i\sum_{j=1}^{n}\sqrt{\tfrac{\beta
            e_{\gamma_{j}}^{2}}{m_{\gamma_{j}} c^{2}}}\int_{0}^{1} d\b\xi_j(s) \cdot
    \A(\r_j+\lambda_{\gamma_j} \b\xi_j (s))\Biggr) \la{B.5}
\end{align}
where
\begin{align}
    V(\r_{i},\b\xi_{i},\r_{j},\b\xi_{j})=\int_{0}^{1}ds \frac{1}
    {|\r_{i}+\lambda_{\gamma_{i}}\b\xi _{i}(s)-\r_{j}-\lambda_{\gamma_{j}}\b\xi _{j}(s)|}
    \la{B.6}
\end{align}
is the Coulomb potential between two wires and the vector potential part a stochastic line
integral that represents the flux of the magnetic field across the wire. The vector potential
is itself a random field distributed by the normalized Gaussian thermal weight $\e^{-\beta
    H_0^\text{rad}}/Z_0^\text{rad}$. Then the partial trace $\avg{\;\;\cdots\;\;}_{{\rm
        rad}}=\frac{1}{Z_{0}^{{\rm rad}}}\Tr_\text{rad}(\e^{-\beta H_0^\text{rad}}\cdots)$ over
the transverse field degrees of freedom in (\ref{force-def}) is easily performed
\begin{align}
    &\avg{\exp\Biggl(i\sum_{j=1}^{n}\sqrt{\tfrac{\beta
            e_{\gamma_{j}}^{2}}{m_{\gamma_{j}} c^{2}}}\int_{0}^{1} d\b\xi_j(s) \cdot
    \A(\r_j+\lambda_{\gamma_j} \b\xi_j (s))\Biggr)}_{{\rm rad}}\nonumber
    \\&=\Bigl(\prod_{i=1}^{n}e^{-\beta e_{\gamma_{i}}^2 W_{m}({\b 0},\b\xi_{i},{\b 0},\b\xi_{i}
        )}\Bigr)\;e^{-\beta\sum_{i<j}^{n}
        e_{\gamma_{i}}e_{\gamma_{j}}W_{m}(\r_{i},\b\xi_{i},\r_{j},\b\xi_{j})}
\la{B.8}
\end{align}
In (\ref{B.8}) $W_{m}$ is a double stochastic integral
\begin{align}
    e_{\gamma_i} e_{\gamma_j} W_{m}(\r_{i},\b\xi_{i},\r_{j},\b\xi_{j}) =
    \frac{1}{\beta\sqrt{m_{\gamma_{i}}m_{\gamma_{j}}}c^{2}} \int \frac{d\bk}{(2\pi)^3}
    \sum_{\mu,\nu=1}^3 j_{\mu}^{*} (\bk, i)G^{\mu\nu}(\bk)j_{\nu}(\bk, j) \la{B.9}
\end{align}
where
\begin{align}
    G^{\mu\nu}(\bk)=\frac{4\pi
        |g(\bk)|^{2}}{|\bk|^{2}}\delta_{tr}^{\mu\nu}(\bk),\quad\delta_{tr}^{\mu\nu}(\bk)=\delta^{\mu\nu}-\frac{k^{\mu}k^{\nu}}{|\bk|^{2}}
    \la{B.10}
\end{align}
is the free field covariance and $\delta_{tr}^{\mu\nu}(\bk)$ the transverse Kronecker
symbol. \footnote{The product in (\ref{B.8}) contains the magnetic self energies of the wires.}
In (\ref{B.9}), ${\b j}(\bk, i)$ is the Fourier transform of the line current ${\b j}({\b
    x},i)=e_{\gamma_i}\int_{0}^{1}d\b\xi_i(s) \delta({\b x}-\r_i-\lambda_{\gamma_i} \b\xi_i
(s))$ associated to the wire $\b\xi$.  One sees that the transverse part of the field gives
rise to an effective pairwise magnetic interaction $W_{m}$ that has (up to a factor) the same
form as the classical energy of a pair of current wires. Its ratio to the Coulomb energy
(\ref{B.6}) is of the order of $k_B T$ divided by the rest mass energy of the particles. It
accounts for orbital diamagnetic effects, which are small in normal conductors. Performing a
small $\bk$ expansion in the integrand of (\ref{B.9}) and noting that $\int_{0}^{1}d\b\xi(s)=0$
one sees that the large distance behaviour of $W_{m}$ is dipolar
\begin{align}
    e_{\gamma_i} e_{\gamma_j} W_{m}(\r_{i},\b\xi_{i},\r_{j},\b\xi_{j}) &\sim
    \frac{1}{\beta\sqrt{m_{\gamma_{i}}m_{\gamma_{j}}}c^{2}}\int_{0}^{1}d\b\xi_{i}(s_{1}) \cdot
    \int_{0}^{1}d\b\xi_{j}(s_{2})\nonumber
    \\&\phantom{\sim}\times \left(e_{\gamma_i}\lambda_{\gamma_j}\b\xi_{i}(s_{1})\cdot
    \nabla_{\r_{i}} \right)\left(
    e_{\gamma_i}\lambda_{\gamma_j}\b\xi_{j}(s_{2})\cdot\nabla_{\r_{j}}\right)\frac{1}{|\r_{i}-\r_{j}|}
    \la{B.11}
\end{align}

Having now identified the basic effective pair interactions between the random wires, namely
the Coulomb potential $V(i,j)$ (\ref{B.6}) and the magnetic potential $W_m(i,j)$ (\ref{B.9}),
it is possible to proceed exactly as in the treatment of classical charged fluids
\cite{Hansen-McDonald}.  One sees that $V(i,j)$ differs from the genuine classical
electrostatic interaction between two charged wires
\begin{align}
    V_{{\rm elec}}(i,j)= \int_{0}^{1}ds_{1}\int_{0}^{1}ds_{2}\frac{1}
    {|\r_{i}+\lambda_{\gamma_{i}}\b\xi _{i}(s_{1})-\r_{j}-\lambda_{\gamma_{j}}\b\xi _{j}(s_{2})|}
    \la{B.14}
\end{align}
by the quantum-mechanical ``equal-time constraint'' imposed by the Feynman-Kac formula. It is
therefore useful to split $ V(i,j)=V_{{\rm elec}}(i,j)+W_{c}(i,j)$, where
\begin{align}
    W_{c}(i,j)=\int_{0}^{1}ds_{1}\int_{0}^{1}ds_{2}(\delta(s_{1}-s_{2})-1)\frac{1}
    {|\r_{i}+\lambda_{\gamma_{i}}\b\xi _{i}(s_{1})-\r_{j}-\lambda_{\gamma_{j}}\b\xi
        _{j}(s_{2})|} \la{B.15}
\end{align}
is the part of $V(i,j)$ due to intrinsic quantum fluctuations ($W_{c}(i,j)$ vanishes if $\hbar$
is set equal to zero). Its large distance behaviour originates from the term bilinear in
$\b\xi_{1}$ and $\b\xi_{2}$ in the multipolar expansion of the Coulomb potential in
(\ref{B.15}). It is dipolar and formally similar to that of two electrical dipoles of sizes
$e_{1}\lambda_{1}\b\xi_{1}$ and $e_{2}\lambda_{2}\b\xi_{2}$.
\begin{align}
    &e_{\gamma_i}e_{\gamma_j}W_{c}(\r_{i},\b\xi_{i},\r_{j},\b\xi_{j})\nonumber
    \\ &\sim \int_{0}^{1}ds_{1}\int_{0}^{1}ds_{2}\,
    (\delta(s_{1}-s_{2}))-1)\left(e_{\gamma_i}\lambda_{\gamma_i}\b\xi_i(s_{1})\cdot \nabla_{\r_{i}}
    \right) \left(
    e_{\gamma_j}\lambda_{\gamma_j}\b\xi_j(s_{2})\cdot\nabla_{\r_{j}}\right)\frac{1}{|\r_{i}-\r_{j}|}
\la{B.16}
\end{align}

Introducing the diagrammatic representation of the correlation functions by Mayer graphs, we perform the usual resummations of $V_{\text{elec}}$-chains to sum the Coulomb divergences. This provides a short range screened potential $\Phi_{{\rm elec}}(i,j)$, as in the classical Debye-H\"uckel mean-field theory. Mayer graphs are reorganized in integrable prototype graphs with bonds 
\begin{align}
&F(i,j) = -\beta e_{\gamma_i} e_{\gamma_j}
    \Phi_{{\rm elec}}(i,j) \label{queq:Fcc}
 \\&\Fr(i,j) = \exp[{-\beta e_{\gamma_i} e_{\gamma_j}
            (\Phi_{{\rm elec}}(i,j)+W_{c}(i,j)+W_{m}(i,j))}]  -1
    +\beta e_{\gamma_i} e_{\gamma_j} \Phi_{{\rm elec}}(i,j)
\label{queq:Fr}
\end{align}
with the constraint of excluded convolution rule between $F(i,j)$ bonds, namely chains of $F$
bonds are forbidden to avoid double counting of the original Mayer graphs.

We now sketch the final steps. To obtain the force, one needs to find the asymptotic form of
the correlation between a wire in $A$ and a wire in $B$. Set $F(i,j)=F_{AB}$ $(F_{AA})$ when
particle i belongs to slab $A$ and particle j belongs to slab $B$ $(A)$, and likewise for
$\Fr(i,j)$. Following the methods of \cite{buenzli-martin}, one shows that the bond $F_{AB}$ is
responsible for the universal term $-\zeta(3)/(8\pi\beta d^3)$ of (\ref{concl}). Some care has
to be exercised with the bond $\Fr_{AB}$ that embodies the effect of field and particle quantum
fluctuations through $W_m$ and $W_c$. It has a dipolar long distance behaviour $F^\text{R}(i,j)
\sim -\beta e_{\gamma_i} e_{\gamma_j}(W_c(i,j) + W_m(i,j)) \sim |\r_i-\r_j|^{-3}$ that might
contribute to the force. In forming the complete correlation function of the two slab system,
the bonds $F_{AB}$ and $\Fr_{AB}$ have to be dressed at their extremities by appropriate
internal correlations of the individual slabs in conformity with the diagrammatic rules. Thus,
the complete expressions that enters in the force formula at large separation are of the form
$G_{AA}\star F_{AB}\star G_{BB}$ and $H_{AA}\star \Fr_{AB}\star H_{BB}$. The formation of the
slabs' internal correlations $G_{AA}$ and $H_{AA}$ in these terms is not the same because of
the excluded convolution rule that applies to $F_{AB}$ but not to $\Fr_{AB}$. Working out the
explicit expressions, one sees that perfect screening sum rules in the system of wires applied
to $G_{AA}\star F_{AB}\star G_{BB}$ imply the universality of the $d^{-3}$ term in
(\ref{concl}), but the term $H_{AA}\star \Fr_{AB}\star H_{BB}$ yields no contribution at order
$d^{-3}$ because of the same sum rules.

Even without going through the detailed calculations, it is clear from the asymptotic forms
(\ref{B.11}), (\ref{B.16}) that the corrections to the electrostatic result
(\ref{force-8pibeta}) due to the quantum nature of the charges and the radiation field are
controlled by the thermal wave lengths $\lambda_\gamma=\hbar \sqrt{\beta/m_\gamma}$, thus small
at high-temperature. Because of the Bohr-van Leeuwen theorem, the free energy
(\ref{free-energy}) of the complete model continuously approaches that of the corresponding
pure electrostatic classical system as the $\lambda_\gamma$ vanish. The force cannot jump by a
factor $2$ in this limit.

One must conclude from this analysis that the discrepancy between (\ref{force-4pibeta}) and
(\ref{force-8pibeta}) is not due to the omission of the transverse part of the electromagnetic
interaction in the classical Coulombic models of refs. \cite{forrester-janco-tellez,
    janco-tellez, buenzli-martin} but should be attribuated to the very fact that fluctuations
inside the conductors are ignored in the calculation leading to (\ref{force-4pibeta}). Hence
(\ref{force-8pibeta}) is the correct asymptotic form of the high-temperature Casimir force. In
other words, the description of conductors by mere macroscopic boundary conditions is
physically inappropriate whenever the effect of thermal fluctuations on the force are
considered.

One the other hand, recent experiments validate the zero temperature formula (\ref{vac}). In
\cite{Bressi} the authors find an experimental agreement with the value of Casimir force's
strength $\pi^2 \hbar c/240$ to a 15\% precision level.  This indicates that fluctuations in
conductors are drastically reduced as the temperature tends to zero and possibly have no more
effect on the force at $T=0$. A full understanding of the cross over from the high temperature
regime (\ref{concl}) to the zero temperature case together with the role plaid by matter and
field fluctuations in the conductors is an open problem.

Finally we like to comment on the Lifshitz versus Schwinger method to take the metallic limit
in their theories of forces between macroscopic dielectric bodies. In \cite{Lifshitz} Lifshitz
obtained the high-temperature large-distance ($\alpha \ll 1$) force between two dielectric
slabs having a static dielectric constant $\epsilon$ as
\begin{align}
    f(d)\sim -\frac{1}{16\pi \beta d^{3}}\int_{0} ^{\infty}
    ds\frac{s^{2}}{\Delta^{2}e^{s}-1},\quad \Delta=\frac{\epsilon+1}{\epsilon-1} \la{lifshitz}
\end{align}
which is easily seen to reduce to (\ref{force-8pibeta}) in the perfect conductor limit of
electrostatics $\epsilon\to\infty$. In \cite{Schwinger}, Schwinger \etalii have proposed to
take the limits in the reverse order, \ie the perfect conductor limit is taken first and the
high-temperature large-distance asymptotics afterwards, resulting in the value
(\ref{force-4pibeta}). In the light of the preceding considerations, the Lifshitz procedure is
the right one to recover the high temperature regime for conductors.

\acknowledgments
We thank B. Jancovici for very fruitful discussions on the subject of this work.

\end{document}